\documentclass[%
 reprint,
 amsmath,amssymb,
 aps,
 pra,
]{revtex4-2}

\usepackage{graphicx}
\usepackage{dcolumn}
\usepackage{bm}
\usepackage{amsmath}
\usepackage{physics}
\usepackage{mathtools}
\usepackage{float}
\usepackage[subrefformat=parens]{subcaption}
\usepackage{listings}
\usepackage{here}
\usepackage{siunitx}
\usepackage{multirow}
\usepackage{ulem}

\newcommand{\vect}[1]{\boldsymbol{#1}}

\DeclareRobustCommand{\erase}{\bgroup\markoverwith{\textcolor{red}{\rule[.5ex]{2pt}{0.4pt}}}\ULon}

\usepackage{xcolor}

\begin{document}

\preprint{APS/123-QED}

\title{Quantum state estimation of multi-partite single photon path entanglement via local measurements}

\author{Hikaru Shimizu$^{1}$}
 \email{shimihika2357@keio.jp}
 
\author{Joe Yoshimoto$^{2,3}$}
 \email{joeyoshiii@keio.jp}

\author{Kazufumi Tanji$^{1}$}

\author{Aruto Hosaka$^{4}$}

\author{Junko~Ishi-Hayase$^{2,3}$}

\author{Tomoyuki Horikiri$^{5,6}$}
 
\author{Rikizo Ikuta$^{7,8}$}
 \email{ikuta.rikizo.es@osaka-u.ac.jp}

\author{Masahiro Takeoka$^{1,9}$}
 \email{takeoka@elec.keio.ac.jp}

\affiliation{%
$^{1}$Department of Electronics and Electrical Engineering, Keio University, 3-14-1 Hiyoshi, Kohoku-ku, Yokohama 223-8522, Japan
} 
\affiliation{%
$^{2}$Department of Fundamental Science and Technology, Keio University, 3-14-1 Hiyoshi, Kohoku-ku, Yokohama 223-8522, Japan
} 
\affiliation{%
$^{3}$Center for Spintronics Research Network, Keio University, 3-14-1 Hiyoshi, Kohoku-ku, Yokohama 223-8522, Japan
}
\affiliation{%
$^{4}$Mitsubishi Electric Corporation, Information Technology R\&D Center, Kanagawa 247-8501, Japan
}
\affiliation{%
$^{5}$Yokohama National University, 79-5, Tokiwadai, Hodogaya, Yokohama, Kanagawa 240-8501, Japan
}
\affiliation{%
$^{6}$LQUOM, Inc., 79-5, Tokiwadai, Hodogaya, Yokohama, Kanagawa 240-8501, Japan
}
\affiliation{%
$^{7}$Graduate School of Engineering Science, Osaka University, Toyonaka, Osaka 560-8531, Japan
}
\affiliation{%
$^{8}$Center for Quantum Information and Quantum Biology, Osaka University, Osaka 560-0043, Japan
}
\affiliation{%
$^{9}$National Institute of Information and Communications Technology (NICT), Koganei, Tokyo 184-8795, Japan
}
\date{\today}

\begin{abstract}
Multipartite entanglement plays a critical role in various applications of quantum internet. 
In these applications, the entanglement is usually shared by the distant parties. 
Experimentally, the distributed entanglement should be estimated by only local measurements. 
Furthermore, for network experiments, 
it is desirable to employ measurement techniques that are straightforward to implement.
In this paper, we propose a method to measure arbitrary multipartite single photon path entangled states by only local measurements. By considering practically reasonable assumptions, our method is relatively easy to implement.  We experimentally demonstrate the utility of this method by reconstructing the density matrix of a 3-qubit W-state.


\end{abstract}

\maketitle


\section{Introduction}
Quantum network\cite{Kimble2008,Wehner2018} will connect quantum devices over distant places and open novel applications of quantum technologies. 
Although the standard way of quantum networking is to share bipartite entanglement such as Bell state among the network nodes, in some applications, including distributed quantum computing\cite{Cacciapuoti2020,Rodney2016}, conference key agreement\cite{Chen2007,Proietti2021,Carrara2023}, and network sensing\cite{Komar2014,Gottesman2012}, sharing multipartite entanglement by more than two parties is required.  
However, similar to the bipartite entanglement transmission, 
direct transmission of multipartite entanglement suffers loss of the channels, which implies strict trade-off between the distance and the entanglement distribution rate. 

One way to practically circumvent the direct transmission limit is to use entanglement by the superposition of zero- and one-photon, such as single photon path entanglement. 
It has been known that heralding two lossy single photon path entangled states via single photon interference detection can overcome the rate-loss trade-off of the direct transmission \cite{Campbell2008,Zo2024} and this idea is recently extended to the loss-tolerant distribution of multipartite entanglement \cite{Roga2023,Carrara2023,Shimizu2024}.



Whereas the single photon path entanglement has its advantage in loss-tolerant entanglement distribution over the network, its measurement and state evaluation in real experiments are not straightforward since the state is a superposition of different photon numbers (typically, zero and one). 
A simple way to evaluate the single photon path entanglement is to combine the two modes via beamsplitter\cite{Zo2024,Lago-Rivera2021,Chou2005,Papp2009,Grafe2014}. It transforms the ideal single photon path entanglement into a single photon state and a vacuum and thus one can effectively evaluate the entangled state by evaluating the non-vacuum single-mode output from the beamsplitter. 
This approach, however, cannot be applied in real quantum network scenario where the entanglement is distributed over the network and thus the measurements are restricted to be local ones. 
To evaluate the state via local measurements, one needs to project the state onto superposition of zero and one photons. This has been performed for bipartite entanglement by preparing weak local oscillators\cite{Tan1991,Li2013,Caspar2020,Liu2023}.

In this paper, we extend the work in Refs.~\cite{Tan1991,Li2013,Caspar2020,Liu2023} and propose the local state evaluation of the multipartite single photon path entanglement. 
Our work aims to develop a technically simple local state evaluation with reasonable assumptions on the state to be evaluated. 
We experimentally implement this idea with the tripartite entanglement (three-qubit W-states in photon number basis) and demonstrate its state reconstruction.
As a related work, Ref.~\cite{Caspar2022} generates the same states and applies local measurements to evaluate their entanglement witness. 
While their work intends to directly measure the witness with minimum assumptions on the state, the goal of our work is to reconstruct the state's density matrix with the simplest measurement setting.

The paper is constructed as follows. The detailed theory of our measurement method is described in Sec.~$\mathrm{I}\hspace{-1.2pt}\mathrm{I}$. 
Section $\mathrm{I}\hspace{-1.2pt}\mathrm{I}\hspace{-1.2pt}\mathrm{I}$ describes our experimnetal setup and the results.
We summarize the paper in Sec.~$\mathrm{I}\hspace{-1.2pt}\mathrm{V}$.

\section{Theory}

In this section, we describe the theoretical background of our measurement and quantum state tomography (QST) scheme. 
For completeness, we start from the measurement of bipartite state distributed in two modes case and then extend it to the multipartite case. 

\subsection{Two-mode single-photon state}
\begin{figure}[h]
    \centering
    \includegraphics[width=0.5\textwidth]{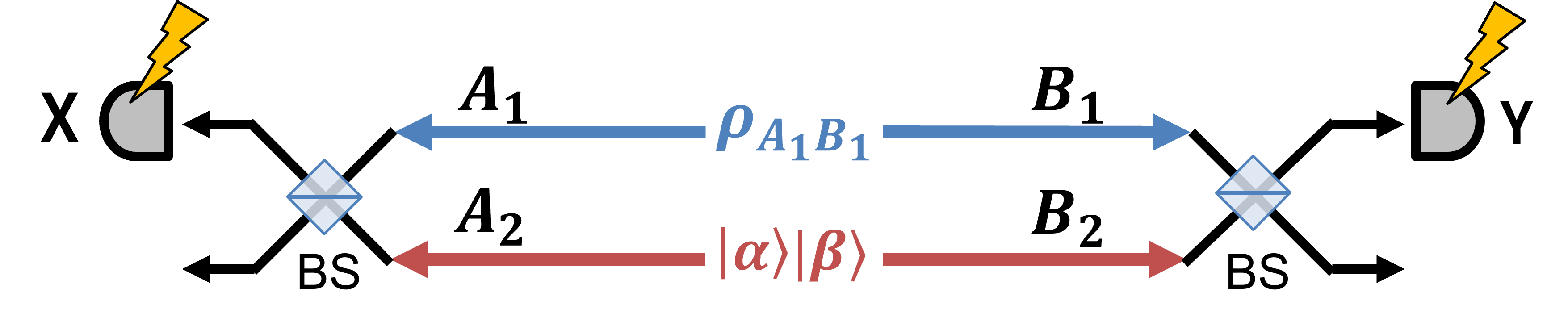}
    \caption{\raggedright 
    Concept of the quantum state tomography of $\rho_{A_1B_1}$ using phase-locked local oscillators $\ket{\alpha}_{A_2}\ket{\beta}_{B_2}$.}
    \label{schematic}
\end{figure}
We start by our QST of the two-mode single-photon state with local measurements, which is illustrated in Fig.~1.   
It basically follows Refs.~\cite{Tan1991,Liu2023}, except that it includes an explicit formula for treating an infinite number of multiphotons in local oscillators (LOs). 
We assume that the density matrix of the two-mode state between modes $A_1$ and $B_1$
are in the space spanned by $\{ \ket{00},\ket{01},\ket{10},\ket{11},\ket{20},\ket{02}\}$, where $\ket{ij}$ is the photon number basis. 
Then the state is represented as 
\begin{equation}
    \label{2mode_arbitrary}
        \rho_{A_1B_1}=
        \begin{bmatrix}
        p_{00} & 0 & 0 & \vect{0} \\
        0 & p_{10} & d & \vect{0} \\
        0 & d^* & p_{01} & \vect{0} \\
        \vect{0} & \vect{0} & \vect{0} & \rho_{2} 
        \end{bmatrix}, 
\end{equation}
where $p_{ij}=\bra{ij}\rho\ket{ij}_{A_1B_1}$ for $i,j=0,1$ is the probability 
where the $i$ and $j$ photons are included in modes $A_1$ and $B_1$, respectively. $\rho_{2}$ represents the unnormalized two-photon state spanned by $\{\ket{11},\ket{20},\ket{02} \}$.
We further assume that $p_{00} \gg p_{01}\sim p_{10} \gg \Tr[\rho_2]$, 
which is usually satisfied in experimental situations \cite{Zo2024,Caspar2020}. 
$p_{ij}$ can be estimated from the single-count measurement 
or coincidence measurement between the two modes. 
In conventional methods, the off-diagonal element $d$ has been estimated by directly mixing photons at the two modes. 
Instead, in our scheme, we prepare a two-mode coherent state  $\ket{\alpha}_{A_2}\ket{\beta}_{B_2}$ as LOs 
where $|\alpha|^2$ and $|\beta|^2$ are comparable with $p_{01}$ and $p_{10}$. 
As shown in Fig.~\ref{schematic}, 
the LOs at modes $A_2$ and $B_2$ are mixed 
with the photons at $A_1$ and $B_1$, respectively, 
by half beamsplitters~(BSs). 
By taking the coincidence counts between modes $X$ and $Y$ after the BSs, 
$d$ is estimated. 

The intuition why our scheme can estimate the off-diagonal term $d$, which corresponds to the nonlocal coherence between modes $A_1$ and $B_1$, by only local measurements is understood as follows. 
When coincidence occurred, with high probability, one photon comes from $\rho_{A_1B_1}$ and the other from the LOs. 
If this event is selected, the pre-measured LOs are of the form $(\ket{01}_{A_2B_2}+\ket{10}_{A_2B_2})/\sqrt{2}$ (where we assume $\alpha = \beta$ for simplicity, but this assumption can be removed easily). 
Then, we can interpret the system as quantum teleportation, where the LOs form the Bell state, the left BS followed by the photon detection at $X$ is the Bell measurement and then the state in mode $A_1$ is teleported to mode $B_2$. 
As a consequence, the right BS effectively mixes two modes in $\rho_{A_1B_1}$. 
That is, the situation is the same as those in Refs.~\cite{Zo2024,Lago-Rivera2021,Chou2005,Papp2009,Grafe2014} and thus the photon detection at mode $Y$ reflects the phase information of $\rho_{A_1B_1}$, i.e. information of $d$.


In the following, we describe the details of the estimation of the off-diagonal element. 
We denote an annihilation operator of mode $M$ 
by $\hat{a}_M$ and the number operator by 
$\hat{N}_{M}=\hat{a}^\dagger_{M}\hat{a}_{M}$. 
We assume the threshold detectors are used in modes $X$ and $Y$. 
The detection efficiencies at the both modes are assumed to be the same for simplicity, 
and much smaller than $1$ 
such that the detection probabilities are proportional to the photon numbers. 
Using transformations of the BSs as 
$\hat{a}_X\rightarrow (\hat{a}_{A_1}+\hat{a}_{A_2})/\sqrt{2}$, 
$\hat{a}_Y\rightarrow (\hat{a}_{B_1}+\hat{a}_{B_2})/\sqrt{2}$ 
and $\rho_{\rm out}\rightarrow \rho_{A_1B_1}\otimes\ketbra{\alpha}_{A_2}\otimes\ketbra{\beta}_{B_2}$, 
where $\rho_{\rm out}$ is the output state of the BSs, 
the coincidence probability between modes $X$ and $Y$ is described by 
\begin{align}
P_{XY}&=\eta^2 \Tr[\hat{N}_X\hat{N}_Y\rho_{\rm out}]\\
    &=P_{\rm NIF} + P_{\rm IF}, 
    \label{coinc_whole}
\end{align}
where $\eta$ is the detection efficiency at each mode. 
$P_{\rm NIF}$ and $P_{\rm IF}$ are 
the non-interferometric and interferometric terms and given as 
\begin{align}
P_{\rm NIF}&=\frac{\eta^2}{4}\left(
    |\alpha|^2|\beta|^2+p_{10}|\beta|^2+p_{01}|\alpha|^2\right),
    \label{eq:NIF}\\
P_{\rm IF}&=\frac{\eta^2}{2}\Re[d\alpha^*\beta]\\ 
&= \frac{\eta^2}{2}|\alpha||\beta||d|\cos{(\theta_d-\phi_\alpha+\phi_\beta)},
\label{coincidence_phase}
\end{align}
where  
$\hat{a}_{A_2}\ket{\alpha}_{A_2}=\alpha\ket{\alpha}_{A_2}$ and 
$\hat{a}_{B_2}\ket{\beta}_{B_2}=\beta\ket{\beta}_{B_2}$ are used. 
$\theta_d$ and $\phi_{\alpha(\beta)}$ are 
the phases of $d$ and $\alpha(\beta)$, respectively.

To estimate $d$, an additional coincidence measurement of $X$ and $Y$ 
without the input of the target state is required. 
The coincidence probability is then $P_{\rm coh} = \eta^2|\alpha|^2|\beta|^2/4$ 
which corresponds to the first term of Eq.~(\ref{eq:NIF}).
For generality, remove the assumption of $\alpha=\beta$ and set their intensity ratio as $|\beta|^2/|\alpha|^2= r$, 
we obtain the visibility $V$ of the interference fringe 
with respect to the change of $\phi_\alpha-\phi_\beta$ as 
\begin{equation}
    \label{v}
    V = \frac{2 \sqrt{r} |d|}{rp_{10}+p_{01}}, 
\end{equation}
where note that the offset $P_{\rm coh}$ is subtracted.  
Then, by using the experimentally observed values of $p_{10}$, $p_{01}$, $r$, and $V$, 
one can estimate $|d|$. 
In addition, by setting $\phi_\alpha-\phi_\beta$ to be $0$ and $\pi/2$, 
the real and imaginary parts of $d$ are obtained, respectively. 


\subsection{Multi-mode single-photon state}
Next, we describe the estimation method to reconstruct the density matrix of an arbitrary single-photon state distributed to three or more modes, 
which is the main part of the theory in this paper. 
We consider a three-mode single-photon state described by 
\begin{equation}
    \label{3mode}
    \begin{matrix}
        \rho_{ABC} = 
    \end{matrix}
    \begin{matrix}
        \begin{bmatrix}
            p_{000} & 0 & 0 & 0 & \vect{0} \\
            0 & p_{100} & d_{AB} & d_{AC} & \vect{0}\\
            0 & d_{AB}^* & p_{010}& d_{BC} & \vect{0} \\
            0 & d_{AC}^*& d_{BC}^* & p_{001} & \vect{0}\\
            \vect{0} & \vect{0} & \vect{0} & \vect{0} & \rho_2
        \end{bmatrix}, 
    \end{matrix}
\end{equation}
where $d_{ij}$ shows the phase coherence between modes $i$ and $j$. 
$\rho_2$ represents the unnormalized two-photon state spanned by $\{ \ket{110}, \ket{101},\ket{011},\ket{200},\ket{020},\ket{002} \}$. 
We assume that $\Tr[\rho_2]$ is much smaller than $p_{100}$, $p_{010}$ and $p_{001}$ (note that this can be easily confirmed in the experiment as shown later). 
Each of the diagonal elements $p_{001}, p_{010}$ and $p_{100}$ is directly obtained from single-count measurement at each mode. 
$p_{000}$ is estimated by $1-(p_{001}+p_{010}+p_{100})$. 

To estimate the off-diagonal element $d_{AB}$, we trace out mode $C$. 
The remaining unnormalized two-mode state between modes $A$ and $B$ becomes 
\begin{equation}
    \label{2mode_traceout}
    \begin{matrix}
        \rho_{AB} = 
    \end{matrix}
    \begin{matrix}
        \begin{bmatrix}
            p_{00}' & 0 & 0 & \vect{0}      \\
            0 & p_{100} & d_{AB} & \vect{0}  \\
            0 & d_{AB}^* & p_{010} & \vect{0}\\
            \vect{0} & \vect{0} & \vect{0} & \rho_{2}' 
        \end{bmatrix},
    \end{matrix}
\end{equation}
where 
$\rho_2'$ is the unnormalized submatrix in the space spanned by $\{ \ket{11}, \ket{20},\ket{02}\}$.
$p_{00}'= p_{000}+p_{001}+p_{002}\sim  p_{000}+p_{001}$ is the probability of vacuum in modes $A$ and $B$. 
The density matrix in Eq.~(\ref{2mode_traceout}) is exactly the same form 
as Eq.~(\ref{2mode_arbitrary}). 
Thus, by applying the same method discussed in the previous section, $d_{AB}$ is estimated. 
Similarly, $d_{BC}$ and $d_{AC}$ are estimated, 
resulting in the successful reconstruction of the three-mode single-photon state 
in Eq.~(\ref{3mode}). 

The reconstruction of an $N$-mode single-photon state is achieved in a similar manner 
to the case of the three-mode state described above. 
By tracing the $N-2$ modes other than the two of interest from the target state, 
the off-diagonal element of the two-mode state can be estimated, 
as long as the two-photon components 
whose contributions are enhanced by $(N-2)$ are negligibly small. 
To complete the reconstruction, the estimation of 
all off-diagonal elements of the $_NC_2=N(N-1)/2$ mode pairs is required. 

\subsection{Consideration of coherence between the vacuum and the single photon}
In the previous sections, we assumed that there is no phase coherence 
between the subspaces spanned by the different photon numbers as in Eq.~(\ref{2mode_arbitrary}). 
In this section, we relax the constraints on the density matrix as 
\begin{equation}
    \label{include_vacuum}
        \rho_{AB}=
        \begin{bmatrix}
        p_{00} & d_{A} & d_{B} & \vect{0} \\
        d_{A}^* & p_{10} & d_{AB} & \vect{0} \\
        d_{B}^* & d_{AB}^* & p_{01} & \vect{0} \\
        \vect{0} & \vect{0} & \vect{0} & \rho_{2} 
        \end{bmatrix}, 
\end{equation}
which includes the non-zero coherence 
between the vacuum and the single photon at each mode. 
While the terms including more than two photons are small enough, 
coherence $d_{A}$ and $d_{B}$ are observed in the coincidence measurement since
 $p_{00}\gg p_{01}, p_{10}$. 
This affects the precise state estimation for the subspace spanned by $\{\ket{01}, \ket{10}\}$. 
Thus, considering the state estimation of Eq.~(\ref{include_vacuum}) is important. 

Starting from Eq.~(\ref{include_vacuum}), 
the interferometric term $P_{\rm IF}$ in the coincidence probability is described by 
\begin{align}
    \label{IF_with_vac}
        P_{\rm IF} = \frac{\eta^2}{2}|\alpha|&|\beta|(|d_{B}||\alpha|\cos{(\theta_{d_{B}}+\phi_\beta)}\\
        &+|d_{A}||\beta|\cos{(\theta_{d_{A}}+\phi_\alpha)}\\
        &+|d_{AB}|\cos{(\theta_{d_{AB}}-\phi_\alpha+\phi_\beta)}), 
\end{align}
which shows the estimation of $d_{A}$ and $d_{B}$ are required to obtain $d_{AB}$. 
The probability $P_{X(Y)}$ of the single counts of mode $X(Y)$ 
depends only on the LO from mode $A_2~(B_2)$ as follows: 
\begin{align}
    \label{single}
        &P_{X} = \frac{\eta}{2}(|\alpha|^2 + p_{10} + 2|\alpha||d_{A}|\cos({\theta_{d_{A}}-\phi_{\alpha}})),\\
        &P_{Y} = \frac{\eta}{2}(|\beta|^2 + p_{01} + 2|\beta||d_{B}|\cos({\theta_{d_{B}}-\phi_{\beta}})),
\end{align}
where $\theta_{d_{A(B)}}$ is the phase of $d_{A(B)}$. 
From this equation, we obtain $d_{A(B)}$ including the phase information 
by setting to $\phi_{\alpha(\beta)} = 0$ and $\pi/2$. 
Using the results, $d_{AB}$ is estimated from Eq.~(\ref{IF_with_vac}). 
Extension of the above strategy to the estimation of the multi-mode single-photon states is rather straightforward. 

\section{Experiment}
\subsection{Experimental setup}
\begin{figure*}[t]
\begin{center}
    \centering
            \includegraphics[width=1\textwidth]{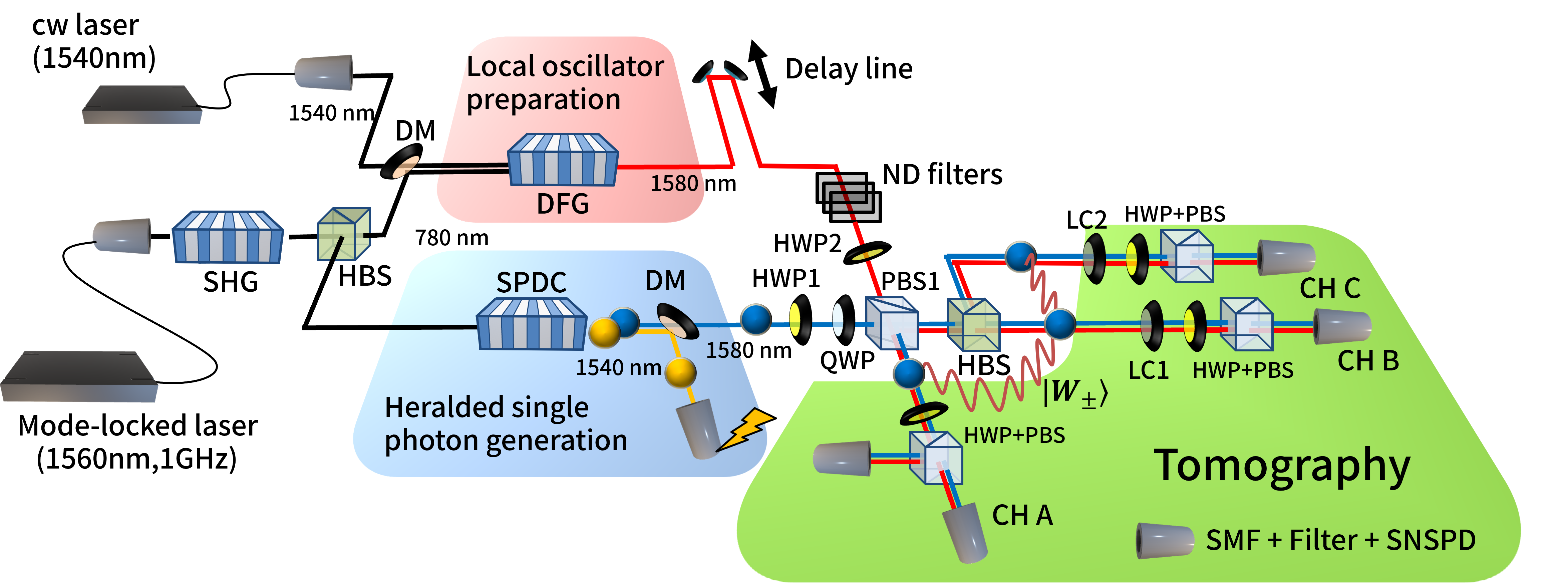}
    \caption{\raggedright The experimental setup. 
    QWP: quater-wave plate, HWP: half-wave plate, 
    DM: dichroic mirror, HBS: half beamsplitter. 
    PBS: polarizing beamsplitter, and  
    LC: liquid crystal. CH A, B, and C are labels for the optical channels. 
    }
    \label{setup}
\end{center}
\end{figure*}
The experimental setup is shown in Fig.~\ref{setup}. 
We use a mode-locked fiber laser with the center wavelength of \SI{1560}{nm}, the repetition rate of \SI{1}{GHz} and pulse width of \SI{4.9}{ps}, 
and a continuous-wave~(cw) laser at \SI{1540}{nm} as light sources. 
The light from the pulsed laser is frequency-doubled by second harmonic generation~(SHG) 
using a type-0 PPLN waveguide. 
The SHG light at \SI{780}{nm} is divided into two optical paths. 
One beam pumps a type-0 PPLN waveguide to prepare a photon pair 
of a signal photon at \SI{1580}{nm} and an idler photon at \SI{1540}{nm} 
via spontaneous parametric downconversion~(SPDC). 
For generating the three-mode single-photon path entanglement, 
the signal photon is heralded by the detection of the idler photon. 
The signal photon in a state of $\alpha\ket{V}+\beta\ket{H}$ adjusted by 
a half-wave plate~(HWP1) and a quarter-wave plate~(QWP) 
is separated into two paths by a polarizing beamsplitter~(PBS1). 
The V-polarized component is reflected to channel A~(CH~A). 
The H-polarized component passing through the PBS1 is further divided 
into two channels of CH~B and CH~C by a HBS. 

Using HWP1 and the QWP before PBS1, 
we prepare two types of the three-mode single-photon states in modes CH~A, CH~B and CH~C as 
\begin{equation}
    \label{eq+++}
    \ket{W_\pm} = \frac{1}{\sqrt{3}}(\ket{100}\pm \ket{010}\pm \ket{001})
\end{equation}
that belong to the class of the W-states, which is a genuine multi-partite entanglement.

The other pulsed beam at \SI{780}{nm} and the cw laser at \SI{1540}{nm} are used 
to prepare LOs at \SI{1580}{nm} 
via difference frequency generation~(DFG) in another type-0 PPLN waveguide 
for measuring the multipartite entangled state. 
After the intensity of the DFG light is attenuated to the single-photon level 
by neutral density~(ND) filters, 
the DFG light is combined with the optical path of the signal photon 
at the PBS1 and is distributed to CH~A, CH~B, and CH~C with the polarization orthogonal to the signal photon. 

We set the intensities of the LOs to be the same in all channels. 

At each channel, 
after the polarization modes of the signal photon and the LO 
are mixed using a HWP and a PBS for estimating the off-diagonal components of the single-photon state 
as shown in Fig.~\ref{schematic}, the photon is detected. 
Relative phases between the signal photon and LO can be changed by 
liquid crystal retarders~(LCs) inserted into optical paths of CH~B and CH~C. 

Each of the photons is coupled to a single-mode fiber~(SMF) and spectrally filtered with the bandwidth of \SI{0.4}{nm} before the detection.
All of the detectors used in the experiment are superconducting nanowire single-photon detectors~(SNSPDs). 
The electric signals from the SNSPDs are used as stop signals 
of the time-to-digital converter~(TDC) 
triggered by the click of the heralding channel as a start signal. 
We collect the two/three/four-fold coincidence events 
between the electric signals coming from one/two/three stop channels and the heralding channel
in the time window of \SI{320}{ps}. 

\subsection{Experimental results}
\begin{figure}[t]
    \centering
    \includegraphics[width=0.4\textwidth]{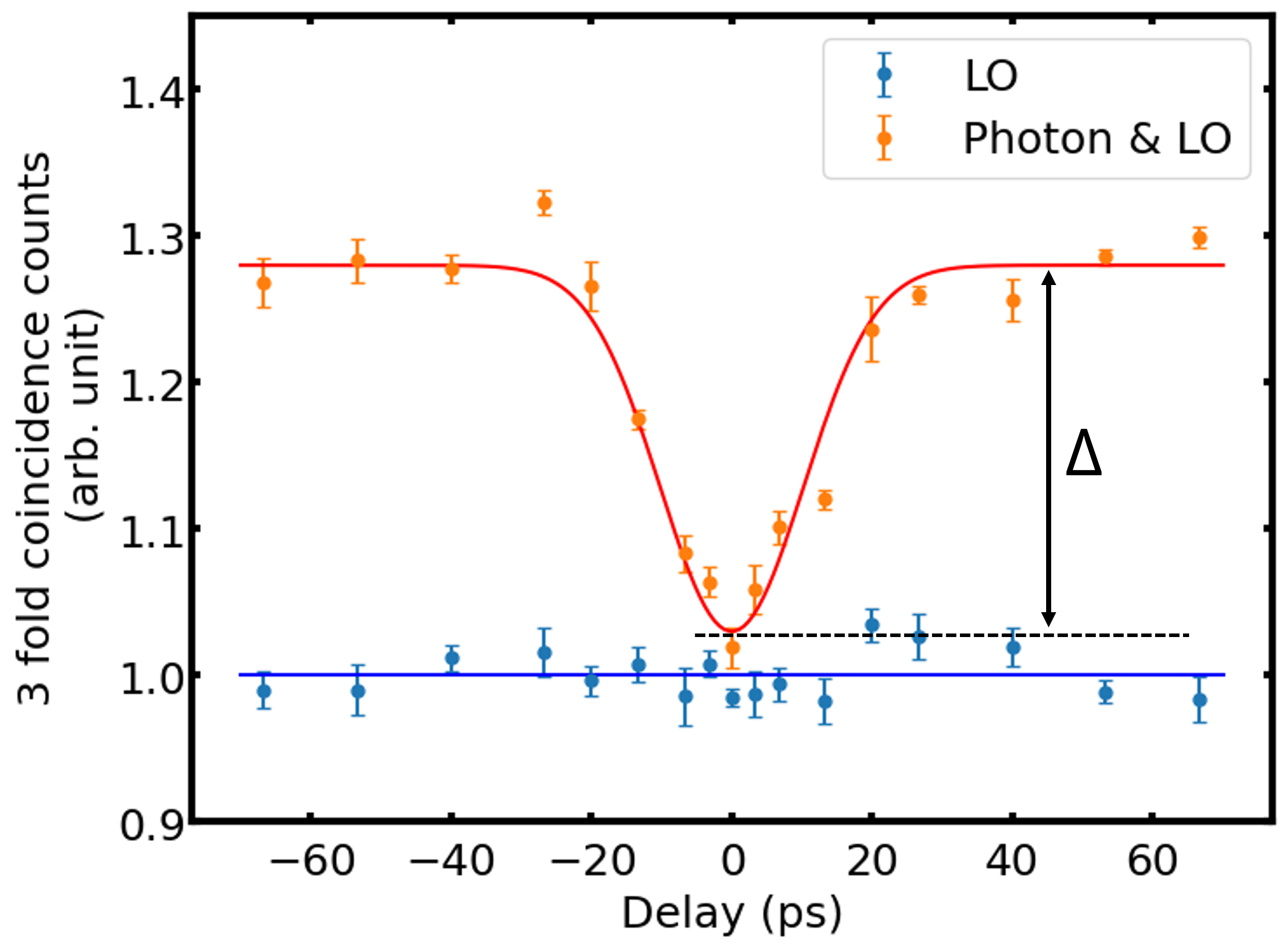}
    \caption{\raggedright 
    Hong-Ou-Mandel interference between LO and heralded signal photon. 
    The counts~(orange) are normalized by 
    those obtained by using only LO without the heralded single photon~(blue). 
   }
    \label{modematch}
\end{figure}

As a preliminary experiment before the QST, 
we observed the Hong-Ou-Mandel~(HOM) interference 
between the heralded signal photon and the LO 
to verify the overlap of the temporal and spectral modes between them. 
It was performed at CH~A by mixing the V-polarized signal photon and H-polarized LO using the HWP and the PBS followed by the two detectors at both output ports. 
The time difference between them was changed by a delay line in the optical path of LO. 
The experimental result is shown in Fig.~\ref{modematch}. 
The vertical axis shows the coincidence count 
normalized by the coincidence count observed by only the multi-photon contribution of the LO without the signal-photon input. 
From Ref.~\cite{Tsujimoto2023}, 
the normalized coincidence count is described by 
\begin{equation}
    \label{HOM}
    N_{\rm cc}(M) = 1 + 2n_{\rm ph}(1-M),
\end{equation}
where $n_{\rm ph}$ is the ratio of the average photon number of the heralded single photon to that of the LO. 
$M$ is the mode overlap probability between the photon and LO. 
Denoting the mode overlap at the HOM dip by $M_{\rm dip}$, we obtain 
\begin{equation}
    \label{dip}
    \frac{\Delta}
    {N_{\rm cc}(0)-N_{\rm cc}(1)} = M_{\rm dip}, 
\end{equation}
where $\Delta = N_{\rm cc}(0)-N_{\rm cc}(M_{\rm dip})$ is the depth of the dip. 
From Fig.~\ref{modematch}, we obtain 
$N_{\rm cc}(0)-N_{\rm cc}(1)=0.28$ and $\Delta=0.25$, 
leading to $M_{\rm dip} = 0.25/0.28 \approx 89.3 \%$.
From the result, we confirmed the photons and LO have a mode overlap 
enough to observe the interference fringe based on the coincidence measurement. 

\begin{figure}[t]
    \centering
    \includegraphics[width=0.45\textwidth]{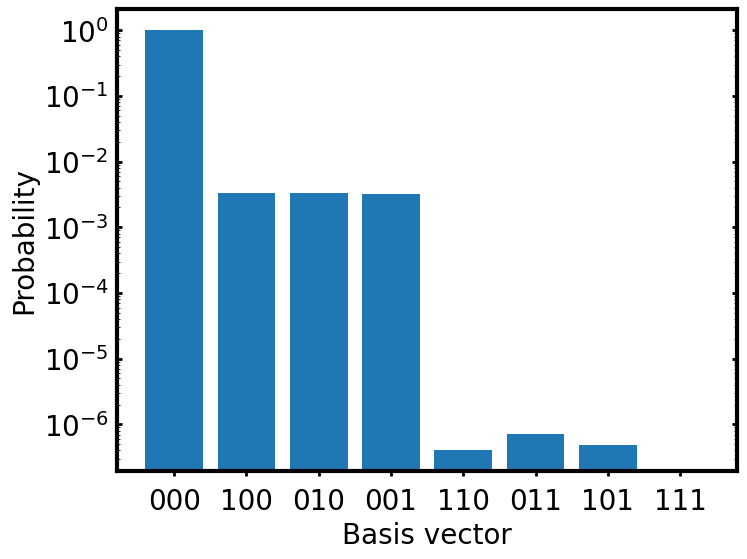}
    \caption{\raggedright Observed probabilities of each photon-number state. }
    \label{diagonal}
\end{figure}

For the QST of the prepared W states 
in Eq.~(\ref{eq+++}),
we first evaluate the diagonal elements of the density matrix 
using two-fold, three-fold and four-fold coincidence counts among CH~A, Ch~B, Ch~C and the heralding channel, without the input of LOs. 
The result is shown in Fig.~\ref{diagonal}. 
The vacuum component is dominant, and the probabilities of states including more than two photons are sufficiently small. 
From the results and the fact that the heralded single photon would have no phase information between the states formed by the different photon numbers, 
the assumption of Eq.~(\ref{3mode}) is well established in the experiment. 

\begin{figure*}[!t]
\begin{center}
    \centering
    \includegraphics[width=1\textwidth]{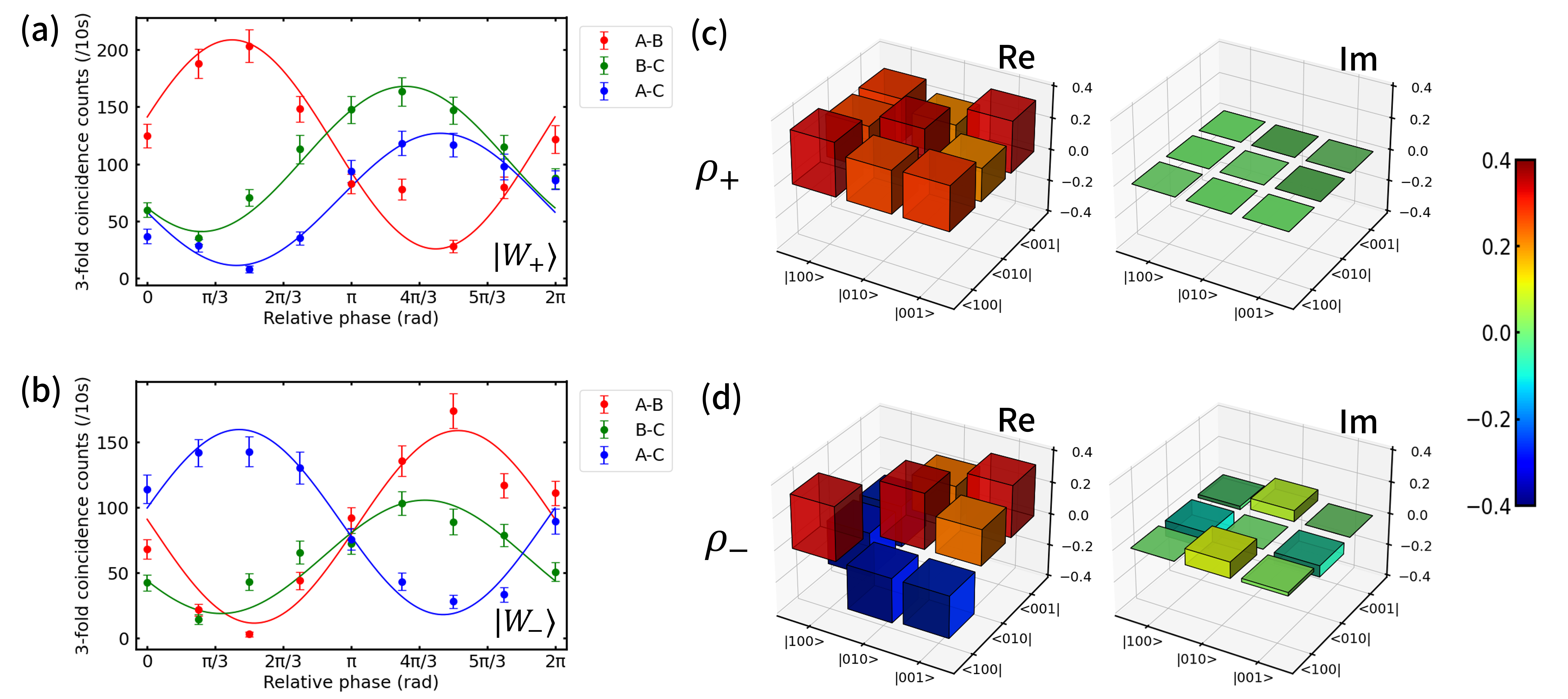}
    \caption{\raggedright 
    (a) Dependencies of the 3-fold coincidence counts on the phases of LOs.  
    (b) Dependencies of the 3-fold coincidence counts on the phases of LOs 
    when the phases of the prepared signal photons in CH B and CH C are flipped from the setting of (a). 
    (c),(d) The reconstructed density matrices $\rho_\pm$. 
    }
    \label{tomography}
\end{center}
\end{figure*}

In the measurement of off-diagonal elements $d_{AB}$ and $d_{BC}$, 
we swept the phase of the LO relative to the target photon 
using LC1. 
For estimating $d_{AC}$, 
the phase of the LO was swept by LC2. 
We collected the three-fold coincidence counts among the heralding channel and two different stop channels with and without the heralded signal photon in each phase. 
By subtracting the latter counts from the former ones, 
we obtained interference fringe as shown in Figs.~\ref{tomography}~(a) and (b).
From the results and Eq.~(\ref{v}), 
the absolute values of $d_{AB}$, $d_{BC}$ and $d_{AC}$ are estimated. 
We regarded the phases of $d_{AB}$ and $d_{BC}$ for $\ket{W_+}$ as zero. 
The relative phases of the other off-diagonal elements were determined 
using the experimental results in Figs.~\ref{tomography}~(a) and (b) consistently. 
From the estimation and Fig.~\ref{diagonal}, 
we reconstructed the density matrices $\rho_\pm$ related to $\ket{W_\pm}$
as shown in Figs.~\ref{tomography}~(c) and (d). 
The fidelities determined by 
$F_\pm = \bra{W_\pm}\rho_{\pm}\ket{W_\pm}$ are 
$F_+= 0.828 \pm 0.017$ and $F_- = 0.864 \pm 0.019$, respectively. 
We summarize the measurement results in TABLE~\ref{table}. 


\begin{table}[t]
  \centering
  \begin{tabular}{c c c c c} 
    \ \ State \ \  & \ \ $ij$ \ \   & \ \ $|d_{ij}|$ \ \  & \ \ $\theta (\rm rad / \pi)$ \ \ & \ \ Fidelity \ \ \rule[-2pt]{0pt}{10pt}\\ \hline\hline
    \multirow{3}{*}{$\ket{W+}$}\rule[-2pt]{0pt}{15pt} & AB & $0.260\pm 0.014$ & $0.000\pm 0.018$ \rule[-2pt]{0pt}{15pt} & \multirow{3}{*}{$0.828\pm0.017 $}\rule[-2pt]{0pt}{15pt}\\ \cline{2-4}
    & BC & $0.203\pm 0.015$ & $0.000\pm 0.030$ \rule[-2pt]{0pt}{15pt} & \rule[-2pt]{0pt}{15pt}\\ \cline{2-4}
    & AC & $0.280\pm0.014$ & $0.000\pm 0.021$ \rule[-2pt]{0pt}{15pt} & \rule[-2pt]{0pt}{15pt}\\ \hline
    \multirow{3}{*}{$\ket{W-}$}\rule[-2pt]{0pt}{15pt}& AB & $0.285\pm0.006$ & $1.044\pm0.017$ \rule[-2pt]{0pt}{15pt} & \multirow{3}{*}{$0.864\pm0.019 $}\rule[-2pt]{0pt}{15pt} \\ \cline{2-4}
    \rule[-2pt]{0pt}{15pt}& BC & $0.246\pm0.020$ & $0.057\pm0.033$ \rule[-2pt]{0pt}{15pt} & \rule[-2pt]{0pt}{15pt} \\ \cline{2-4}
    \rule[-2pt]{0pt}{15pt}& AC & $0.265\pm0.017$ & $0.969\pm0.019$ \rule[-2pt]{0pt}{15pt} & \rule[-2pt]{0pt}{15pt}\\ \hline
  \end{tabular}
\caption{Summary of the experimental results.}
  \label{table}
\end{table}

The estimated fidelity might appear worse than the intrinsic fidelity due to the following possible measurement errors. 
One of the errors is the mode mismatch between the signal photon and the LO. 
When the probability of the mode matching for mode $A(B)$ is denoted by $V_{A(B)}$, 
the interferometric term in Eq.~(\ref{coincidence_phase}) becomes 
$P_{\rm IF}\rightarrow \sqrt{V_AV_B}P_{\rm IF}$. 
When all mode overlaps are $M_{\rm dip}$, 
the fidelities of the states to $\ket{W_+}$ and $\ket{W_-}$ will be estimated to be 0.888 and 0.927, respectively. 
We guess regarding the other errors, 
the fluctuations of the relative phases among the LOs would be induced 
because we did not perform any phase stabilization of the optical paths 
during the measurement, which took about one hour.


\section{Conclusion}
In conclusion, we propose a method to estimate a multi-mode single-photon path entanglement by expanding the method that applies weak-field homodyne measurement. 
In the proposed method, the phase correlation of the photon-number states is estimated by the two-photon interference between the two-mode LO and the two-mode single-photon state, which are the corresponding subsystems of the multi-mode states. 
We theoretically showed that the off-diagonal elements of the density matrix of the multi-mode single-photon state is estimated from the interference fringes of the two-photon interferences while sweeping relative phases of LOs. Using the diagonal elements estimated from the single count measurement without the LOs, the density matrix is reconstructed. 
We experimentally demonstrate the applicability of our method by reconstructing the density matrices of the three-mode single-photon states in the W-state class with the high fidelities. 
We believe that our method will be useful as a standard method for estimating multipartite single-photon states shared among distant parties. 


\newpage
\begin{acknowledgments}
T.H., R.I., and M.T. acknowledge the members of the Quantum Internet Task Force for the comprehensive and interdisciplinary discussions on the quantum internet.
This work was supported by Program for the Advancement of Next Generation Research Projects, Keio University; JST CRONOS, JPMJCS24N6; JST Moonshot R\&D, JPMJMS226C, JPMJMS2066; R \& D of ICT Priority
Technology Project JPMI00316; FOREST Program, JST JPMJFR222V.
\end{acknowledgments}


\begin{thebibliography}{}

\bibitem{Kimble2008}H. J. Kimble. The quantum internet. Nature {\bf453}, 1023 (2008).

\bibitem{Wehner2018}S. Wehner, D. Elkouss, and R. Hanson. Quantum internet: A vision for the road ahead. Science {\bf362}, 6412 (2018).

\bibitem{Cacciapuoti2020} A. S. Cacciapuoti, M. Caleffi, F. Tafuri, F. S. Cataliotti, S. Gherardini, and G. Bianchi. Quantum internet: Networking challenges in distributed quantum computing. IEEE Network {\bf34}, 137-143 (2020).

\bibitem{Rodney2016}R. V. Meter and S. J. Devitt. The path to scalable distributed quantum computing. Computer {\bf49}, 31-42 (2016).

\bibitem{Chen2007}K. Chen, H.-K. Lo, Multi-partite quantum cryptographic protocols with noisy GHZ states. Quantum Inf. Comput. {\bf 7}, 689–715 (2007).

\bibitem{Proietti2021}M. Proietti, J. Ho, F. Grasselli, P. Barrow, M. Malik, and A. Fedrizzi. Experimental quantum conference key agreement. Sci. Adv., {\bf7}, 23 (2021).

\bibitem{Carrara2023}G. Carrara, G. Murta, and F. Grasselli. Overcoming fundamental bounds on quantum conference key agreement. Phys. Rev. Applied 19, 064017 (2023)

\bibitem{Komar2014}P. Kómár, E. M. Kessler, M. Bishof, L. Jiang, A. S. Sørensen, J. Ye, and M. D. Lukin. A quantum network of clocks. Nature Physics {\bf10}, 582-587 (2014).

\bibitem{Gottesman2012}D. Gottesman, T. Jennewein, and S. Croke. Longer-baseline telescopes using quantum repeaters. Phys. Rev. Lett. {\bf109}, 070503 (2012).

\bibitem{Campbell2008}E. T. Campbell and S. C. Benjamin. Measurement-Based Entanglement under Conditions of Extreme Photon Loss. Phys. Rev. Lett. {\bf 101}, 130502 (2008).

\bibitem{Zo2024}W. Zo, B. Bilash, D. Lee, Y. Kim, H. Lim, K. Oh, S. M. Assad, Y. Kim. Entanglement swapping via lossy channels using photon-number-encoded states. arXiv:2405.03951 (2024).

\bibitem{Roga2023}W. Roga, R. Ikuta, T. Horikiri, and M. Takeoka. Efficient dicke state generation in a network of lossy channels. Phys. Rev. A {\bf108}, 012612 (2023).

\bibitem{Shimizu2024}H. Shimizu, W. Roga, D. Elkouss, and M. Takeoka. Simple loss tolerant protocol for GHZ-state distribution in a quantum network, arXiv:2404.19458 (2024).




\bibitem{Lago-Rivera2021}D. Lago-Rivera, S. Grandi, J. V. Rakonjac, A. Seri, and H. de Riedmatten. Telecom-heralded entanglement between multimodes solid-state quantum memories. Nature {\bf 594}, 37-40 (2021).

\bibitem{Chou2005}C. W. Chou, H. de Riedmatten, D. Felinto, S. V. Polyakov, S. J. van Enk, and H. J. Kimble. Measurement-induced entanglement for excitation stored in remote atomic ensembles. Nature {\bf 438}, 828-832 (2005).

\bibitem{Papp2009}S. B. Papp, K. S. Choi, H. Deng, P. Lougovski, S. J. van Enk, and H. J. Kimble. Characterization of Multipartite Entanglement for One Photon Shared among Four Optical Modes. Science {\bf 324}, 5928 (2009).

\bibitem{Grafe2014}M. Gräfe, R. Heilmann, A. Perez-Leija, R. Keil, F. Dreisow, M. Heinrich, H. Moya-Cessa, S. Nolte, D. N. Christodoulides, and A. Szameit. On-chip generation of high-order single-photon W-states. Nature Photonics {\bf 8}, 791–795 (2014).

\bibitem{Caspar2020}P. Caspar, E. Verbanis, E. Oudot, N. Maring, F. Samara, M. Caloz, M. Perrenoud, P. Sekatski, A. Martin, N. Sangouard, H. Zbinden, and R. T. Thew. Heralded Distribution of Single-Photon Path Entanglement. Phys. Rev. Lett. {\bf 125}, 110506 (2020).

\bibitem{Liu2023}J. -L. Liu, X. -Y. Luo, Y. Yu, C. -Y. Wang, B. Wang, Y. Hu, J. Li, M. -Y. Zheng, B. Yao, Z. Yan, D. Teng, J. -W. Jiang, X. -B. Liu, X. -P. Xie, J. Zhang, Q. -H. Mao, X. Jiang, Q. Zhang, X. -H. Bao, and J. -W. Pan. Creation of memory-memory entanglement in a metropolitan quantum network. Nature {\bf 629}, 579-585 (2024).

\bibitem{Tan1991}S. M. Tan, D. F. Walls, and M. J. Collett. Nonlocality of a single photon. Phys. Rev. Lett. {\bf 66}, 252 (1991).

\bibitem{Li2013} L. Li, Y. O. Dudin, and A. Kuzmich. Entanglement between light and an optical atomic exitation. Nature {\bf 498}, 466 (2013).

\bibitem{Caspar2022}P. Caspar, E. Oudot, P. Sekatski, N. Maring, A. Martin, N. Sangouard, H. Zbinden, and R. Thew. Local and scalable detection of genuine multipartite single-photon path entanglement. Quantum {\bf 6}, 671 (2022).

\bibitem{Tsujimoto2023} Y. Tsujimoto, R. Ikuta, K. Wakui, T. Kobayashi, and M. Fujiwara. 
Quantum State Tomography of Qudits via Hong-Ou-Mandel Interference. 
Phys. Rev. Appl. {\bf 19}, 014008 (2023).
    
\end{thebibliography}
\end{document}